# A Mathematical Formalization of Self-Determining Agency

Yoshiyuki Ohmura*, Earnest Kota Carr, Yasuo Kuniyoshi

The University of Tokyo

## Abstract

Defining agency is an extremely important challenge for cognitive science and artificial intelligence. Physics generally describes mechanical happenings, but there remains an unbridgeable gap between these and the acts of agents. To discuss the morality and responsibility of agents, it is necessary to model acts; whether such responsible acts can be fully explained by physical determinism remains an ongoing debate. Although we have already proposed a physical "agent determinism" model that appears to go beyond mere mechanical happenings, we have not yet established a strict mathematical formalism to eliminate ambiguity. Here, we explain why a physical system can follow coarse-graining agent-level determination without violating physical laws by formulating supervenient causation. Generally, supervenience including coarse graining does not change without a change in its lower base; therefore, a single supervenience alone cannot define supervenient causation. We define supervenient causation as the causal efficacy from the supervenience level to its lower base level. Although an algebraic expression composed of the multiple supervenient functions does supervenes on the base, an index sequence that determines the algebraic expression does not supervene on the base. Therefore, the sequence can possess unique dynamical laws that are independent of the lower base level. This independent dynamics creates the possibility for temporally preceding changes at the supervenience level to cause changes at the lower base level. Such a dual-laws system is considered useful for modeling self-determining agents such as humans.

## 1. Introduction

Defining agency is critical for cognitive science and artificial intelligence. According to McGregor (2016), it is necessary to explain the difference between two types of events: (i) An act, which is performed by an agent, (ii) A mere happening, which occurs as a result of purely mechanical reasons. In light of remarkable advancements in artificial intelligence, it has become an urgent task to establish a framework for understanding agency and discussing the responsibility to non-living systems. However, it is unclear whether acts can be fully explained by physical determinism.

Barandiaran et al. (2009) characterized agency by individuality, interactional asymmetry, and normativity. In terms of individuality, if an external agent is required to define the difference between the system and the surrounding environment, then an infinite regress occurs. Therefore, they considered that the condition for individuality was to define its own identity as an individual, distinguishing itself from its surroundings. It is true that if we consider movement as a change in the boundary between a system and its environment, then the infinite regress problem of the initiation of an act can also be resolved by a system that self-determines its boundary. However, it is possible to consider the problem of infinite regress in the initiation of an act as distinct from the issue of boundary self-determination. This is because the problem of initiation of an act can be resolved when the cause of motion is generated internally within the system rather than triggered externally from it, and fulfilling this condition requires the distinction between the system and its environment as a prerequisite. We address how to define an initiation mechanism of actions within a system, rather the boundary self-determination between the system and its environment. Interactional asymmetry appears to be particularly important for explaining the initiation of an act. To explain the initiation of an act, we must clarify the asymmetric causal mechanism of internal intervention that does not directly interfere with other parts of the system without a causal transmission. The reason it is internal intervention rather than external intervention is that external intervention cannot resolve an infinite regress of the initiation of an act. Finally, normativity refers to the agent system itself determining goals or norms.

McGregor (2016) criticized this definition as phenomenological and difficult to mathematically formalize, proposing instead an approach that interprets systems from an external observational perspective using the intentional stance (Dennett, 1991) without explicitly modeling the mechanism of initiation of an act. Subsequently, many approaches have modeled agents externally (McGregor, 2017; McGregor & Chrisley, 2020; Kenton et al., 2023; Biehl & Virgo, 2023; Virgo et al., 2025). However, predicting data using the intentional stance is distinct from elucidating causal mechanisms (Kaplan, Craver, 2011). Identifying mechanisms is a primary goal of theory construction. Information about mechanisms is important for causal explanations. (Woodward, 2011). Approaches that predict behavior without acknowledging the causal efficacy of mental states such as belief or desire cannot provide a scientific explanation of the causal mechanisms underlying mental processes. Furthermore, the interpretation of agency through these approaches may vary depending on the level of agency assumed by the observer. It is well known that phenomena that cannot be explained by an assumed theory tend to be ignored (Brewer, Lambert, 2001). A theory of agency should aim to be a theory that covers all of the following levels and should not ignore phenomena that cannot be explained by existing theories.

We propose that agency can be classified into the following levels: 1. Goal-directedness 2. Goal-switching 3. Goal-generating 4. Agent-determinism. Which level qualifies as 'agency' may differ among researchers, but most intentional stance approaches can be interpreted as moving from level 1 toward level 4. In contrast, philosophical discussions of morality and free will typically assume level 4 agency.

In philosophy, when physical determinism or randomness is claimed to be incompatible with free will and responsibility, agent causation (Chisholm, 1982) is often invoked. However, debates continue over whether agent causation is fundamental (O'Conner, 2009) or can be explained by emergent causation (Steward, 2012; Mayer, 2018), leaving unclear whether agent determination is compatible with physicalism or naturalism. Consequently, level 4 agency has not been formally defined in the field of science. However, we believe that the agent-determinism approach is more promising than the intentional stance approach for explaining the aforementioned three characteristics of agency, because agent determinism is precisely the argument that attempts to explain the initiation of action.

Recently, we argued that although systems are physically implemented, their behavior does not necessarily adhere to single-level laws alone (Ohmura & Kuniyoshi, 2025). The introduction of agent-specific laws at the coarse-grained level creates room for agent determinism. Single-level interactions between an agent and the environment cannot account for the initiation of an action because they lead to either a dissatisfying model of circular causation or an irresolvable situation of infinite regression. Neither provides a reason for the initiation of the action. Therefore, inter-level causation is a promising model for explaining agent determinism. However, our previous study did not provide a mathematical formalization, leaving unclear how laws independent of the laws of the component parts can be introduced at the coarse-grained level without violating physical laws.

Ohmura and Kuniyoshi (2025) proposed that the supervenient coarse-graining macro level can possess unique dynamical laws that are underdetermined by the component parts level laws through self-referential feedback control mechanism. However, their proposed mechanism is limited to the relationship between the whole and its parts and can be discussed as a causal relationship within the more general supervenient-subvenient relationship. This study aims to mathematically formulate supervenient causation and clarify the conditions under which a supervenience level possesses causal efficacy over the lower base level (subvenience level). The apparent contradiction between the following two points must be resolved: 1) supervenience remaining unchanged without changes in the subvenience base, 2) a cause at the supervenience level that must not directly change the effect at the subvenience level. We demonstrate that the supervenience level, which is composed by the multiple supervenient-subvenient relationships, must have independent laws. Finally, we propose a model of self-determining agency.

## 2. About Supervenient Causation

One proposed method for naturalizing agent determinism is to employ causation from supervenience to subvenience (the lower base), referred to as supervenient causation. The relationship between supervenience and subvenience is defined as follows: a change in supervenience cannot occur without a corresponding change in subvenience (Kim, 1998). Originally introduced in moral philosophy and later adopted by Davidson (1970), this terminology has become widely used in the philosophy of mind. While a change in subvenience does not necessarily entail a change in supervenience (i.e. multiple realizability), any change in supervenience is accompanied by a change in subvenience. Thus, supervenience can be regarded as a function of subvenience. Sperry's notion of downward causation, understood as whole-to-parts causation, is included within the scope of supervenient causation (Sperry, 1991).

Since mental phenomena cannot occur without changes in neural activity, they are considered to supervene on physical phenomena. Consequently, causal efficacy from supervenience to subvenience has been expected to account for the influence of mental phenomena on neural activity—namely, mental causation or agent causation (Searl, 1980; Sperry, 1991; Steward, 2012; Mayr,2018).

In humans, behavior arises from the activity of muscles, and its causes can be traced back to neural activity in the motor neurons and, preceding that, to neural activity in the motor cortex. How far should such impending causal regression continue? It is difficult for us to establish clear criteria for determining which neural activity should be regarded as the initiator of the action. Circular causation does not exist solely within the brain; since all physical phenomena potentially contain circular causation, we cannot believe that circular causation explains the initiator of the action. Furthermore, within the complex interaction network, it is impossible to identify the definite initiator of an act. The simplest way to explain the intuition that the initiator of an act in the brain is solely the "I" is to consider that a function corresponding to the role of the "I" exists within the brain model and possesses causal efficacy. However, as long as we consider causality at a single level, we fall into infinite homunculi.

Consequently, the idea emerged that the agent that initiates an act cannot be explained by the single level causation alone and inter-level causation is essential to explain agent-determinism. From this perspective, the two main approaches in the philosophy of mind—event causation and agent causation—discuss the relationship with inter-level causation.

Event causation is the view that mental events such as desires and beliefs cause actions (Davidson, 1963; Mayr, 2018), and agent causation is the view that the agent itself is the cause of actions (Chisholm, 1982; O'Conner, 2009; Steward, 2012). Regarding the former, if mental events such as desires and beliefs are the causes of actions, then the agent or "I" becomes unnecessary (Goldman, 1970; Himmelreich, 2024). As for the latter, the problem lies in the unclear mechanism by which the agent exerts causal influence on its environment. To naturalize the agent causation, Popper and Mitchell (2022) proposed that holistic integration is needed. Holistic integration holds that a system so deeply interconnected that it does not make sense to decompose it into its component parts because the true essence of the system exists in the relations between those parts. However, such holistic integration is insufficient to explain the causal efficacy beyond the interacting parts level.

Agent causation often assumes causal efficacy from the whole to its parts to resolve the problem of the unclear mechanism of agent causation (Steward, 2012; Steward, 2017). Whole-to-parts causation is promising as an explanation of agent causation because it is intrinsic macro causation, but it can also be used to explain mental event causation. Davidson (1970) assumes that the causation of mental events can be explained by supervenient causation. Searle (1983) also assumes supervenient causation in explaining intentional causation (i.e. mental event causation). Thus, opinions vary on whether supervenient causation constitutes mental event causation or agent causation. In this paper, we resolve these confusions.

Agent causation and mental event causation are not mutually exclusive. For example, Clarke (1993) claims that agent causation is necessary for acting with free will, but not for acting per se, thereby modifying Chisholm's agent causation. In recent years, the integration of agent causation and mental event causation has been discussed (O'Conner and Churchill, 2004; Himmelreich, 2024; Martinez, 2024).

Since 1996, Clarke has proposed a different integration method than the one used in 1993 (Clarke, 1996; Clarke, 2003). Our integration of agent causation and event causation is similar to that of Clarke (1993). The most important similarity lies in the view that the difference between the two kinds of causation is a difference in the type of related entities, not the relation. Figure.1 shows our integration of agent causation and event causation. Agent causation involves the temporal changes of discrete states such as beliefs or desires, while event causation involves feedback control of physical states to satisfy desired states. Changes in discrete states are caused by supervenience level dynamics, which we call supervenient causes. And this dynamics needs prior events, such as experiential representation. We assume that the supervenient level dynamics corresponds to the role of "I".

Here, beliefs and desires must be represented by supervenient entities because modifying or configuring these elements within the system requires well-defined variables and functions. It is important to note that the feedback control targets specified at the supervenience level are discrete in nature and therefore cannot directly specify the absolute value of the physical quantities at the subvenience level (e.g., joint angles, or head position).

The absolute value of a physical quantity is often used as a target in feedback control. However, unless we assume an objective scale, absolute values are meaningless. Objective scales are formed by social consensus and cannot be self-determined. Therefore, we must focus on relative relationships, rather than the absolute value of a physical quantity. We consider actions not as behaviors directed toward target values in a state space, but as behaviors that realize a certain relative relationship.

The change in physical states at the subvenience level brought about by a change in the discrete states is supervenient causation, corresponding to mental event causation. A causal transmission in the supervenient causation is achieved through inter-level feedback control mechanisms (Ohmura, Kuniyoshi, 2025).

When considering agents as causes of an act, like Chisholm (1982), it becomes difficult to model how agents determine the act based on past or present experiences. In our model, “I” plays a role in causing changes in the discrete states of the supervenience level based on current and past experiences. Because independent dynamics, “I”, exists at the supervenient level, the initiation of an act can be explained by a causal transmission from a supervenient cause to changes in the subvenience level. Because the causal transmission depends on feedback control, whether the subvenience level is affected by the supervenience cause can be determined by the system itself. For example, during sleep, the initiation of an act is restricted. Furthermore, convergence to the feedback control target set at the supervenience level establishes a steady state. When the target changes, movement resumes, which explains the initiation of an act. Consequently, the system’s behavior cannot be explained by the dynamics at subvenience level alone in our model.

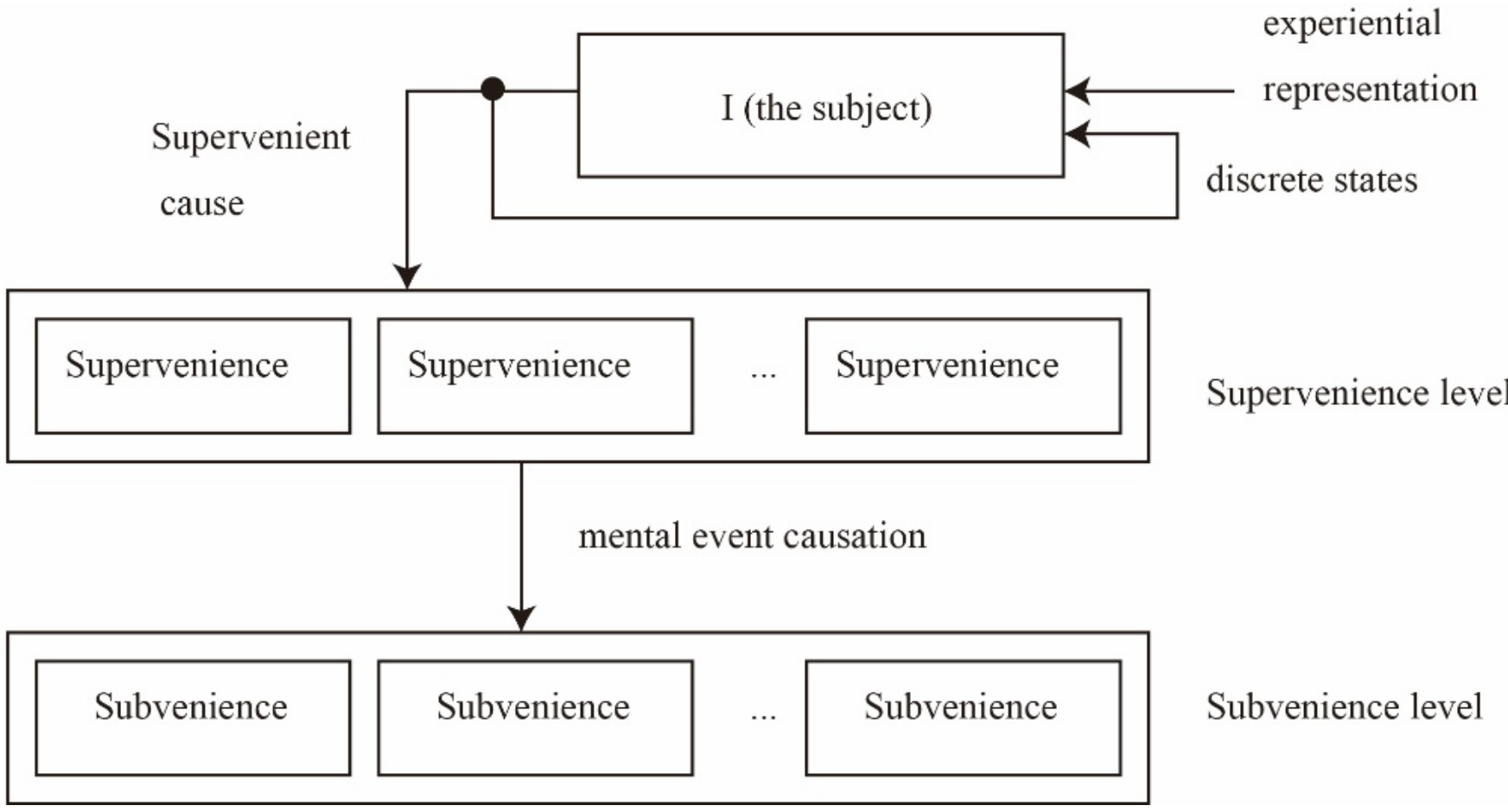

Figure.1: The Correspondence Between Our Model and the Debate in the Philosophy of Mind. A role of “I” changes the discrete states such as beliefs and desires based on experience. This corresponds to agent causation. Changes in discrete states correspond to supervenient causes. However, these causes at the supervenience level do not directly affect the subvenience level. A causal transmission mechanism from the supervenient cause to subvenience level is required. This supervenient causation corresponds to mental event causation in a philosophy of mind.

# 3. Concerning the definition of causality

To explain the initiation of an act, it is necessary to clarify the asymmetric causal mechanism with a supervenient cause that does not interfere directly with the subvenience level without a causal transmission. In this study, we assume that a supervenient cause temporally precedes changes in subvenience level. Merely treating causations at the supervenience level as boundary conditions or constraints is insufficient to explain a supervenient cause; yet little discussion has addressed how supervenient causes can temporally precede their effects (Craver, Bechtel, 2007; Baumgartner, Casini, 2017). This study argues that defining a supervenient cause provides a strong motivation for introducing independent laws at the supervenience level.

Although various definitions of causation exist, discussions in the philosophy of mind often employ token causation rather than type causation (Davidson, 1967; Davidson, 1970; Kim, 1998; Gallow, 2022). While type causation is widely used when addressing scientific knowledge (Halpern, 2016), token causation is more appropriate for debates concerning agent causation and the responsibility for actions, since these issues involve causal relations in particular events. This study focuses on causal mechanisms (Salmon, 1985; Woodward, 2011) to examine causality in particular events.

Inter-level causation is often criticized as not being able to be called causation (Craver, Bechtel, 2017). When defining causation by manipulability, intervention at the supervenience level must not directly affect the subvenience level without a causal transmission (Pearl, 1991; Woodward, 2003). Woodward (2015) proposes treating causation from supervenience as a causal relationship by relaxing the intervention condition, but such relaxation solely to enable supervenient causation is non-essential (Baumgartner, Casini, 2017). To define supervenient causation, we must clarify the intervention method that does not directly change the effect at the subvenience level. In this study, we show that independent intervention at the supervenience level can be defined without relaxing the intervention condition. In conventional inter-level causality debates (Sperry, 1991; Steward, 2012; Hoel et al., 2013; Rosas et al. 2020), a supervenient cause capable of independent intervention is not clearly defined. Our position is that supervenient causation without a supervenient cause is not actual causation.

Interventionist account of causation, which defines causation by manipulability (Pearl, 1991; Menzies and Price, 1993; Mathias, 2023), is unsuitable for testing the causation from the supervenience to the subvenience. Because changes in supervenience cannot occur independently of subvenience, intervention on supervenience alone is impossible (Baumgartner, 2009; Shapiro, 2010; Raatikainen, 2010; Kistler, 2013; Baetu, 2021; Kinney, 2023; Hoffmann-Kolss, 2024). The definition of causation based on manipulability suffers from the problem that it cannot distinguish between asymmetric causal relationships and symmetric equivalence relations (Woodward, 2003). To distinguish them, a supervenient cause must be clarified and must temporally precede the effect at the subvenience level. For example, when analyzing the causal relationship between neural spikes and their coarse-graining (e.g. EEG), the interventionist approach cannot be used because the changes in the supervenient coarse-graining cannot occur without the changes in neural activity.

An interventionist account of causation should not be employed without clarifying an intervention method at the supervenience level that does not directly affect the subvenience level. Furthermore, to apply the interventionist account of causality, we must clarify the causal mechanism from the supervenience level to the subvenience level (Salmon, 1985). Without this clarification, there is a risk of mistaking non-causal relationships for causal ones when using an interventionist account of causation for inter-level causation.

In addition, it is necessary to avoid conflating information with causation (Sánchez-Cañizares, 2023). According to Pearl, causal relation cannot be expressed in the language of probability (Pearl, 2010), implying that no data-driven method can clarify causation. Information-theoretic approaches, extracting information structures from biological data analysis (Hoel et al., 2013; Rosas et al., 2020; Mediano et al., 2025), are not essential for defining supervenient causation without clarifying the causal mechanism. Therefore, we must clarify the logical conditions under which a supervenient cause does not interfere with the subvenient base level without a causal transmission.

We define causality as a pair consisting of a cause and a causal transmission mechanism. A causal transmission mechanism is the mechanism that conveys the cause event to the effect event. The cause event must not affect the effect event without this causal transmission mechanism. When a supervenient cause does not satisfy this condition, we do not consider it a causal relationship. This is because our goal is to clarify an asymmetric causal mechanism between levels.

Causation from supervenience to subvenience has often been criticized for leading to epiphenomenalism (Kim, 1998; Craver and Bechtel, 2007; Robinson, 2010). Epiphenomenalism asserts that supervenience merely accompanies subvenience without possessing causal efficacy (Huxley, 1896; Broad, 1925). For example, if mental events such as beliefs or desires supervene on neural activity, it is impossible to manipulate mental events independently of changes in neural activity. The central question is whether mental events, which do not vary independently, can be granted causal efficacy beyond neural activity. Dennett questions the very definition of epiphenomenalism (Dennett, 1991), redefining the term does not resolve the problem. Accordingly, this study employs the definition commonly used in the philosophy of mind. And this study mathematically formalizes epiphenomenalism to reveal methods for avoiding it.

Causation is temporally asymmetric: the causes precede the effects, and they never occur simultaneously. From this perspective, a change in supervenience cannot serve as the cause of a change in subvenience when considering only one supervenience-subvenience relation. Therefore, this study defines a supervenience level as a set composed of multiple supervenient entities and a subvenience level as the corresponding set of subvenient entities. We have previously proposed a mechanism by which the supervenient whole level exerts causal efficacy on its own parts (supervenience level), termed a self-referential feedback control mechanism (Ohmura & Kuniyoshi, 2025). We assume that multiple mathematical functions (neural networks X and Y) at the supervenience level, the functions supervene on synaptic weights and neural states (subvenient entities x and y). Then, the equations for defining the algebraic structure can be used to define the feedback error at the supervenience level. For example, a feedback error to satisfy commutativity (XY = YX) can be used to adjust the subvenient entities x and y. We define the negative feedback control of subvenient entities to reduce the feedback error as a self-referential feedback control mechanism. In this case, the supervenience level and subvenience level share the same physical entities, we can call this mechanisms whole-to-parts causation. In this study, we mathematically formalize supervenient causation—a general form of whole-to-parts causation—to clarify the conditions under which it becomes possible. We argue that changes in algebraic equations at the supervenience level constitute supervenient causes capable of independent intervention.

# 4. Preparation

The supervenience-subvenience relationship can generally be regarded as a function mapping subvenience to supervenience. Subvenience consists of observable physical quantities and can be represented as vectors. In this study, we define subvenience as $\mathrm{SUB}_i \subset \mathbb{R}^{n_i}$ for an index set $i \in I$, $I \subset \mathbb{N}$. The entire set of subvenience is given by:

$$\mathrm{SUB} \coloneqq \bigcup \mathrm{SUB}_i.$$

We define supervenience as a function whose domain is direct product of $N$ sets $\mathbb{R}^m \times \mathbb{R}^m \times \cdots \times \mathbb{R}^m$ and codomain is $\mathbb{R}^m$, as follows:

$$\mathrm{SUP[N]}: \mathbb{R}^m \times \mathbb{R}^m \times \cdots \times \mathbb{R}^m \to \mathbb{R}^m.$$

For $X^i, X^j \in \mathrm{SUP[1]}$, we assume that a composition operation $X^i \circ X^j$ exists for some $i, j \in I$.

Supervenience is often modeled as a vector, such as coarse-grained states (Israeli, 2006; Hoel et al., 2013; Flack, 2017). In our definition, this can be viewed as a special case where the composition operation is addition and the domain is $\mathbb{R}^m$. Since supervenience only requires that it does not change without a change in subvenience, it need not be restricted to variables; we therefore define it more generally as a function.

Since supervenience is generally determined by subvenience alone (Yoshimi, 2007), the following bridge function $B$ exists:

$$B: \mathrm{SUB} \to \mathrm{SUP[N]}.$$

In this study, for upper case letters $X^i \in \mathrm{SUP}[\mathrm{N}_i]$ as supervenience, lower case letters $x^i \in \mathrm{SUB}_i$ as its corresponding subvenience, and $b_i \in B$ as the bridge law, we define:

$$X^i = b_i(x^i).$$

Let arrays $\boldsymbol{x} \coloneqq (x^0, x^1, \dots)$, $\boldsymbol{X} \coloneqq (X^0, X^1, \dots)$, $b \coloneqq (b_0, b_1, \dots)$, and represent the bridge law as:

$$\boldsymbol{X} = b(\boldsymbol{x}).$$

It is important to note that the subvenience determining the supervenience is not causation, but constitution (Craver, Bechtel, 2007).

All elements $v \notin \mathrm{SUB}$ are considered to belong to a set of un-subvenience USUB. The elements of USUB do not play an essential role in the following formulation.

# 5. Formulation: Epiphenomenalism

When the dynamics of subvenience $\boldsymbol{x}$ and supervenience $\boldsymbol{X}$ can be expressed as:

$$\boldsymbol{x}_{t+1}, v_{t+1} = p(\boldsymbol{x}_t, v_t), \boldsymbol{X}_{\boldsymbol{t}} = b(\boldsymbol{x}_{\boldsymbol{t}}),$$

we define the supervenience $\boldsymbol{X}$ as an epiphenomenon. This is because any changes in supervenience occurs simultaneously with changes in subvenience and is uniquely determined by the bridge law. Since the above dynamics are unaffected by any cause at the supervenience level, supervenience level lacks causal efficacy.

Even if statistically, using supervenience yields higher predictive performance, this does not constitute evidence that supervenience possesses causal efficacy. Furthermore, as mentioned earlier, employing an interventional account of causation for causal inference is also inappropriate to clarify the supervenient causation because intervention at supervenience level must not interfere with the subvenience level without a causal transmission. Therefore, rather than defining epiphenomenalism by an information theoretical method, it is necessary to clarify precisely which models fall into epiphenomenalism.

Defining epiphenomenalism in terms of such a general physical system may provoke criticism. However, the original proposal of epiphenomenalism was formulated under the strong influence of physical determinism (Huxley, 1874; Broad, 1925; Campbell, 2001), making this connection almost inevitable. Dennett (1991) argued that if epiphenomenalism were correct, we could not meaningfully speak about mental properties; thus, he criticized the philosophical notion of epiphenomenalism as overly strong and even absurd—a problem commonly referred to as the self-stultification objection (Robinson, 2006). Dennett interprets epiphenomena as side effects accompanying functional processes, but this definition lacks practical relevance for discussions of supervenient causation. This is because, in debates on the initiation of an act or free will, a supervenient cause must temporally precede changes in subvenience, whereas side effects, such as shadows caused by light, cannot clarify the definite initiator of an act, like circular causation.

Moreover, Dennett (1991) claims similarity between his position and Huxley's original definition of epiphenomenalism, however this positioning has faced valid criticism (Campbell, 2001). Therefore, rather than conveniently redefining epiphenomenalism as Dennett does, we adopt the meaning traditionally employed in the philosophy of mind.

**Lemma 1**

If the dynamics of subvenience can be expressed as:

$$\boldsymbol{x}_{t+1}, v_{t+1} = p(\boldsymbol{x}_t, v_t, \boldsymbol{X}_t(\boldsymbol{d}_t)), \boldsymbol{X}_t = b(\boldsymbol{x}_t),$$

then $\boldsymbol{X}$ is an epiphenomenon. Here, input $\boldsymbol{d}$ is an array of $d \in \mathbb{R}^m \subset$ USUB.

This follows immediately from:

$$\boldsymbol{x}_{t+1}, v_{t+1} = p\big(\boldsymbol{x}_t, v_t, \boldsymbol{X}_t(\boldsymbol{d}_t)\big) = p\big(\boldsymbol{x}_t, v_t, b(\boldsymbol{x}_t)(\boldsymbol{d}_t)\big) = p'(\boldsymbol{x}_t, v_t).$$

Essentially, the supervenience level becomes unnecessary for describing the dynamics of the subvenience level. This implies that supervenience is causally impotent. Furthermore, in this model, intervention at the supervenience level can never be performed without intervention at the subvenience level. Since supervenient causes cannot be defined, supervenience levels cannot possess causal efficacy. To distinguish causality from symmetric relations, it is important to identify the supervenient cause. This is the reason for epiphenomenalism.

Biological systems contain mechanisms that are influenced by the holistic integration of factors such as pH levels, temperature, osmotic pressure, and electrical waves (Sourjik, Wingreen, 2012; Lundqvist et al., 2018; Di Talia. Vergassola, 2022). The existence of such a mechanism appears to contradict this Lemma at first glance. However, when discussing the causal mechanisms of agent systems, implementation differences that do not produce functional differences are not essential. This lemma does not claim that mechanisms influenced by coarse-grained physical quantities do not exist in living organisms. We believe that these coarse-grained physical quantities, lacking autonomy and failing to account for supervenient causation, do not affect the epiphenomenon debate. We assume that differences at the hardware implementation level are not essential for defining agency.

## 6. Formulation: Supervenient Causation

We defined a self-referential feedback control mechanism as an inter-level feedback control that drives changes in feedback error—caused by state transitions at the supervenience level—to converge to zero through negative feedback control at the subvenience level (Ohmura & Kuniyoshi, 2025). In this study, we provide a formalization of supervenient causation as a generalization of this mechanism and clarify why supervenient causation is possible.

Consider an index sequence $c = [i_0, i_1, \dots ] \in C$, where $i_k \in I \subset \mathbb{N}$. Each index corresponds one-to-one with an element of SUP. Let $s: C \to E$ be a mapping from sequences to mathematical expressions.

For example:

$$s([i_k]) = X^{i_k},$$

$$s(c) = s([i_0, (, i_1, i_2, )]) = X^{i_0}(X^{i_1}, X^{i_2}),$$

where $X^{i_0}: \mathbb{R}^m \times \mathbb{R}^m \to \mathbb{R}^m \subset \mathrm{SUP}[2]$ and $X^{i_1}, X^{i_2}: \mathbb{R}^m \to \mathbb{R}^m \subset \mathrm{SUP}[1]$. Thus, $s(c): \mathbb{R}^m \to \mathbb{R}^m$. We assume that for any sequence $c$ the domain and codomain of mathematical expression $s(c)$ are $\mathbb{R}^m$.

By rearranging the index sequence, diverse mathematical expressions can be generated using supervenient functions. This mechanism has an added benefit: it allows novelty to be generated from within the system, ensuring that novelty is not limited to coming from outside the system.

Supervenience $X^i$ does not change without changes in its subvenient $x^i$. To define supervenient causation, we must introduce the intervention method at the supervenience level that does not directly affect the subvenience level. To achieve this, we utilize the assumption that algebraic expressions composed of multiple supervenient functions can be modified independently of subvenience.

**Supervenient Cause**

Let $c_0^L, c_0^R, c_1^L, c_1^R, \cdots \in C$, and we define an array of index sequence pairs:

$$\boldsymbol{c}^{pair} := [(c_0^L, c_0^R), (c_1^L, c_1^R), \dots ].$$

Each index sequence pair defines the right-hand and left-hand sides of an equation that defines feedback error. This equation is represented by $s(c_k^L) = s(c_k^R)$.

The following shows an example of commutativity: $X^{i_1}X^{i_2} = X^{i_2}X^{i_1}$.

$$s(c^L) = s([i_1, i_2]) = X^{i_1}X^{i_2},$$

$$s(c^R) = s([i_2, i_1]) = X^{i_2}X^{i_1}.$$

We introduce the following temporal law: for $w_T, w_{T+1} \in$ USUB,

$$\boldsymbol{c}_{T+1}^{pair}, w_{T+1} = P(\boldsymbol{c}_T^{pair}, w_T).$$

Since $\boldsymbol{c}_T^{pair}, w_T$ are not supervenience, they can change independently of subvenience. We define the state transition of $\boldsymbol{c}^{pair}$ as a supervenient cause. The $\boldsymbol{c}^{pair}$ can be intervened without affecting the subvenience level.

Let:

$$\left[\left(s(c_0^L), s(c_0^R)\right), \left(s(c_1^L), s(c_1^R)\right), \ldots\ \right] := [(e_0^L, e_0^R), (e_1^L, e_1^R), \ldots\ ],$$

where expressions $e_0^L, e_0^R, e_1^L, e_1^R, \cdots \in E$ are defined by supervenience and composition operations. Thus, a supervenient cause determines the dynamics of these expressions.

**Feedback Error**

Define feedback error $er_k \in \mathbb{R}^m$ using $d \in \mathbb{R}^m \cap$ USUB as:

$$er_k(d_k) \coloneqq [s(c_k^L) - s(c_k^R)](d_k) = [e_k^L - e_k^R](d_k)$$

Feedback error is determined by the operator defined by the difference between the left-hand and right-hand sides and the input $\boldsymbol{d} = (d_0, d_1, \ldots)$.

Let:

$$\boldsymbol{er} = (er_0, er_1, \ldots) \coloneqq \varepsilon(\boldsymbol{c}^{pair})$$

Consider the influence of this feedback error on subvenience dynamics. For $\boldsymbol{x}_t, \boldsymbol{x}_{t+1} \in$ SUB and $v_t, v_{t+1}, \boldsymbol{d}_t, \boldsymbol{d}_{t+1} \in$ USUB :

$$\boldsymbol{x}_{t+1}, v_{t+1}, \boldsymbol{d}_{t+1} = p(\boldsymbol{x}_t, v_t, \varepsilon(\boldsymbol{c}_t^{pair})(\boldsymbol{d}_t)).$$

Here, changes in subvenience do not alter $c_k^L, c_k^R$, but supervenient functions $e_k^L, e_k^R$ change because they belong to supervenience. Simultaneously, expression $[e_k^L - e_k^R]$ changes according to $c_k^L, c_k^R$. Thus, the feedback error is influenced independently by both changes at subvenience level and index sequence dynamics at supervenience level. Because feedback error can be independently influenced by both subvenience and supervenience level, supervenient causation becomes possible.

**Supervenient Causation**

$$\boldsymbol{x}_{t+1}, v_{t+1}, \boldsymbol{d}_{t+1} = p\left(\boldsymbol{x}_t, v_t, \varepsilon\left(\boldsymbol{c}_t^{pair}\right)(\boldsymbol{d}_t)\right),$$

$$\boldsymbol{c}_{T+1}^{pair}, w_{T+1} = P(\boldsymbol{c}_T^{pair}, w_T).$$

The dynamics at the subvenience level are governed not only by subvenience level laws but also by supervenience level dynamics $P$, forming a dual-laws model. The former is a continuous system because its states are defined by real numbers. The latter is a discrete system, because the index sequences are composed of natural numbers.

For supervenient causes to influence subvenience level, it is necessary to assume an appropriate dynamics $p$. If the subvenience level is controlled to reduce feedback errors, then the equation at the supervenience level can affect the subvenience level. Causal inference may also be possible by intervening on the supervenient cause.

If $P$ does not exist, then $\boldsymbol{c}_{T+1}^{pair} = \boldsymbol{c}_T^{pair}$, and since $\varepsilon\left(\boldsymbol{c}_t^{pair}\right)$ is fully determined by supervenience alone, Lemma 1 implies that supervenience $X$ becomes an epiphenomenon. Therefore, supervenience level dynamics $P$ are necessary to avoid epiphenomenalism. Because $P$ is independent of subvenience level dynamics $p$, supervenient cause affects the subvenience level. In this case, changes in subvenience cannot be described solely by subvenience level laws. Furthermore, if the cardinality of SUP is 1, then er $\equiv 0$, making supervenient causation impossible.

Thus, to avoid epiphenomenalism, the system must obey not only subvenience level laws but also supervenience level laws. The existence of such lawful dynamics at the supervenience level can be termed agent determinism. Agent determinism implies that system behavior is not determined solely by subvenience level laws, yet remains physically implementable without violating physical laws, thereby preserving the assumption of physical causal closure. Previous claims that supervenient causation is impossible stemmed from conflating physical causal closure with intra-level causal closure [Ohmura and Kuniyoshi, 2025]; however, we demonstrate that supervenient causation is theoretically possible and confirm that the supervenient causation is impossible without introducing supervenient level laws.

# 7. A Model of agency

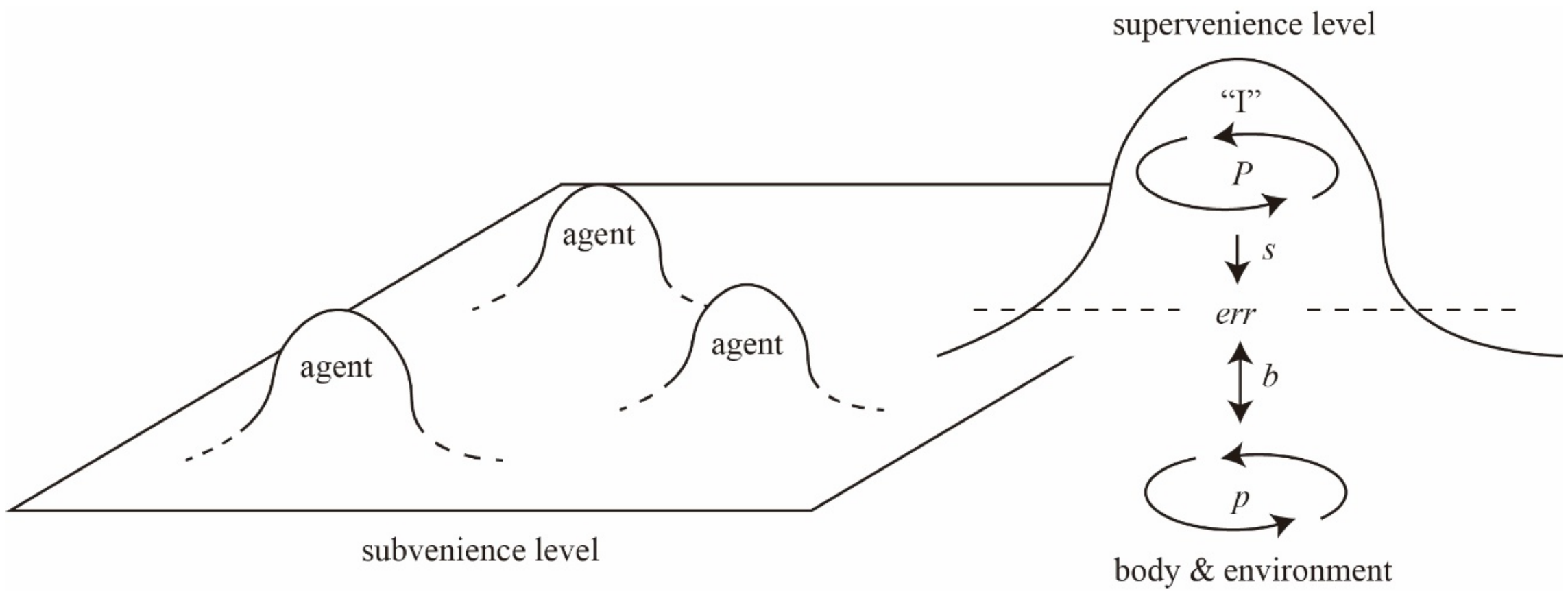

Figure. 2: A model of agency. The dynamical law $P$ that alters the index sequence corresponds to "I". When "I" changes the index sequence, a supervenient cause is triggered. This modifies feedback errors. At the same time, it also changes through the bridge function $b$ by the changes in subvenience. The boundary of agent-specific dynamics does not lie between the body and the environment but rather exists between hierarchical levels within the system. The supervenience level dynamical law characterizes the agency.

Feynman once wrote, "What I cannot create, I do not understand." Scientific explanation should reveal causal mechanisms without relying on black boxes (Salmon, 1971; Machamer et al., 2000; Ross & Bassett, 2024) and the mechanisms should be able to be constructed from mockups. We believe that predictive approaches based on statistical models from an external observational perspective do not lend themselves to developing minimal mechanisms and are thus insufficient for scientific explanations.

Using the formalized concept of supervenient causation, we propose a causal mechanism of agency (Figure 2). We take advantage of the fact that the feedback error $er$ is independently influenced by both changes in subvenience and changes in the index sequence at the supervenience level. Feedback error can be reduced by controlling the subvenience level, because feedback error is composed of multiple supervenient functions and each supervenient function is determined by the corresponding subvenience and is controllable.

Subvenience directly interacts with the body and the environment through the sensors and motors within the agent system. The subvenience level, composed of the agent's body and the environment, can be treated as a unified dynamical system. This level is shared among multiple agents (Figure 2). Each agent possesses its own unique dynamics at its supervenience level.

When focusing on the interaction between the environment and the agent system, it is common to formalize the agent and the environment as Moore machines, each with distinct dynamics (Beer, 1995; Biehl & Virgo, 2023; Virgo et al., 2025; McGregor & Virgo, 2025). Moore machines treat agent systems as a single dynamics, not as a model where coarse-grained level dynamics exist inside the system. In Moore machines, the environment and agent cannot be distinguished in the formulation, exhibiting high symmetry. This makes it difficult to explain the uniqueness of the agent system. Defining the agency should involve elucidating the mechanisms specific to the agent system.

We believe that the Moore machine model is an insufficient model for elucidating agency. There are three reasons. First, the theoretical existence of coarse-grained level dynamics within physical systems demonstrates that single-level dynamics models lack the generality to represent all physical systems. Second, when the model and reality mismatch, it may lead to erroneous conclusions. Third, generally, information structure analysis alone cannot clarify causal mechanisms. Therefore, we believe clarifying the causal mechanisms that explain the aforementioned three characteristics of agency is the most promising methodology.

The agent system is driven not only by external triggers such as sensory input, but also by triggers generated internally. Such triggers are not limited to observable physical quantities. Changes in discrete states at the coarse-grained level can serve as triggers. We argue that possessing triggers from higher coarse-grained level is a defining characteristic of agent systems to explain the initiation of an act.

In our formulation, agent-specific dynamical law is defined within the system, and the boundary of two dynamics does not exist between the environment and the system. While our approach resembles Moore machines in assuming two dynamics, it differs in that we posit distinct dynamics between hierarchical levels. The supervenience level possesses discrete state dynamics, while the subvenience level, including both the agent physical body and the environment, possesses continuous quantity dynamics. This distinction provides the criterion for categorizing them into two distinct types of dynamics. In conventional Moore machines, agents and environments are both reduced to the same type of dynamics, and there is no way to distinguish between them.

The dynamical law $P$ at the supervenience level, corresponding to “I”, modifies the index sequence. This dynamical law gives rise to supervenient causes. These supervenient causes induce changes in the feedback error $err$. When this feedback error is reduced through feedback control at the subvenience level, supervenient causation is established. At this point, while the supervenience level can alter the equation that defines feedback error, this level cannot implement feedback control to continuously reduce that error due to discrete nature. The supervenience level unilaterally determines the mathematical expressions for calculating feedback errors. We argue that the qualitative difference between these two dynamics is the explicit asymmetry required for a satisfactory model of agent causation.

The most important feature of a supervenient cause is that it is intrinsic when the dynamics that modify the index sequence exists within the system. The initiation of an act must be intrinsic because if causal explanations depend on external factors, the problem of an infinite homunculi arises. This is one of the motivations for using supervenient causation to define agent systems.

Barandiaran et al. (2009) proposed three conditions for defining agency: individuality, interactional asymmetry, and normativity. According to Baltieri and Suzuki (2025), none of the theories analyzed in this paper—the free energy principle, integrated information theory, and dynamical systems—meet all the highlighted requirements.

We clarify the causal mechanism within the agent systems to explain the initiation of an act. In our model, the initiation of an act is not determined by external trigger alone and depends on agent-specific coarse-grained level dynamics, so it does not lead to infinite explanatory regress (Barandiaran et al. 2009). Furthermore, in our model, the role of initiator of an act, "I" exists within the system as the supervenient dynamics of index changing the algebraic equations, thereby satisfying the condition of individuality. As discussed earlier, the condition of interactional asymmetry is also met. Supervenient cause does not change subvenience level without a causal transmission and is independent of the subvenience level. This independent supervenient cause triggers the initiation of an act. Regarding the final condition, although it is not yet fully clear how "meaningful" goals are determined internally within the system, it is possible to generate goals for feedback control within the mechanism. Furthermore, since feedback control is often employed to realize goal-directed behavior, the condition of normativity can also be satisfied.

Thus, our model is justified as a proposed definition of agency.

# 8. Discussions

To enable supervenient causation, it is necessary to define a supervenient cause that does not interfere with subvenience level without a causal transmission. However, since changes in supervenience cannot occur without changes in subvenience, this requires assuming a special mechanism. The self-referential feedback mechanism exploits the fact that the index sequence used to define feedback error is not supervenience. This allows the index sequence to have dynamics independent of those at the subvenience level. Because both the index sequence and subvenience level independently influence feedback error, we define changes in the index sequence as a supervenient cause, and the resulting changes in subvenience mediated by feedback error as supervenient causation.

We have defined changes in supervenience as occurring simultaneously with changes in subvenience. In contrast, some have considered the influence from subvenience to supervenience as causal (Dennett, 1991; Shapiro, 2010). For example, Dennett discusses whether shadows are epiphenomena. However, changes occurring at different times generally correspond to physical changes in different spaces. Redefining epiphenomena does not resolve the problem nor explain initiation of an act or agent determinism. Therefore, we constructed our theory based on general definition in the philosophy of mind.

Kim (1991) argued that supervenient causation is impossible. However, Kim conflates physical causal closure with intra-level causal closure, rendering his argument invalid (Mayer, 2018; Ohmura and Kuniyoshi, 2025). Accordingly, we propose that supervenient causation can be realized through a self-referential feedback control mechanism (Ohmura and Kuniyoshi, 2025). The aim of this study was to mathematically formalize this mechanism to clarify the reason why supervenient causation becomes possible.

To define a supervenient cause, it is necessary to introduce changes at the supervenience level that are independent of subvenience. In this study, we clarified that the dynamics of the index sequence, which is not supervenience, can be independent of subvenience-level dynamics. Such index sequences resemble DNA codes but are distinctive in that they modify algebraic expressions at the supervenience level and exert causal efficacy on subvenience via feedback error. This interpretation was not presented in our previous work and constitutes one of the contributions of this study.

The idea that predefined desires and beliefs cause actions (Davidson, 1963) implies that, as long as desires and beliefs merely supervene on neural activity, they lack causal efficacy beyond what neural activity explains. However, if desires and beliefs are mechanized as symbolic indices and structures and formed according to laws at the supervenience level, they can become supervenient causes, alter subvenience, and trigger initiation of actions. In such cases, feedback control at the subvenience level operates to satisfy desires formed at the supervenience level, which can be regarded as an explanation of goal-generation and goal-directedness.

Our model represents a formulation of modified agent causation as proposed by Clarke (1993). Traditional Chisholm-style agent causation assumes that an act occurs because the agent is its cause. By emphasizing the distinction between event causation and agent causation, the latter has often been regarded as a mysterious form of causality that cannot be fully explained. In contrast, Clarke argues that *"the only difference between the two kinds of causation concerns the type of entities related, not the relation."*

In our model, the dynamical laws at the subvenient level govern changes in the neural and physical states, while the dynamical laws at the supervenient level govern changes in the index sequence. Although the entities involved in causation differ across these levels, both can be described as forms of dynamical laws. Furthermore, when changes in the index sequence alter feedback error and thereby influence the subvenient level, this supervenient causation corresponds not to agent causation but to intentional or mental event causation (Davidson, 1963; Searl, 1980). Thus, our model suggests that the supervenience level laws play the role of "I", who is the initiator of acts and the experiencer, and the causal efficacy from supervenience level to subvenience level corresponds to intentional causation to explain the act caused by the desire or belief. We assume that the system that includes the role of “I” is agent.

Dual-process models have been developed widely and successfully in many areas of research (Sloman, 1996; Evans, 2008; Kahneman, 2011). These theories suggest that System 1 is associative, automatic, and fast, and that System 2 is rule-based, deliberate, flexible, and slow. However, the interaction between these two systems remains unclear (Herrmann-Pillath, 2019; Grayot et al. 2024) and dual-process theory has not been considered in philosophy due to its unclear definition (Markus, 2019). We believe our formulation provides the theoretical basis for the dual-process model and the interaction mechanism.

Our model also offers several implications for defining the concept of an agent. Under the traditional *as-if stance*, agents are evaluated externally from an observational perspective, whereas our approach allows explicit modeling of the causal mechanism of the agent. Because it is physically implementable, our model avoids the ontological issues that arise in agent causation. Moreover, while event causation explains actions through reasons for acts and thus renders agents or “I” unnecessary, our model requires agent-specific laws as supervenient-level dynamics. A supervenient cause is a reason for actions and is generated by the dynamics of the supervenient level, which corresponds to the definite initiator of actions within the agent system.

Supervenient causation differs from ordinary causal mechanisms in its characteristics. First, because supervenience cannot intervene independently of subvenience, interventionist approaches cannot be applied to causal inference. Furthermore, in our model, the dynamics of subvenience are not determined solely by the dynamical laws at the subvenience level; they also involve independent dynamical laws at the supervenience level. The behavior of such a system cannot be described as a function of observable state variables alone. Hidden Markov models also cannot be applied to the dual-laws model. Models that do not incorporate dynamical laws at the supervenience level are forced to treat essential aspects of the causal structure as if they were noise (Froese, 2024). Finally, to apply our model to clarify the causal mechanism in the actual brain system, the intervention method in discrete states is unclear because it is highly dependent on implementation. Research methods targeting systems where supervenient causation may exist will become an important research topic in the future.

Formalizing agency is a critical challenge in cognitive science and artificial intelligence. Existing definitions of agency range widely—from those satisfying only goal-directedness to those incorporating morality and responsibility—making it difficult to establish a single definition. With the recent development of AI, definitions of agency applicable to discussions of morality and responsibility are increasingly demanded. However, previous work has lacked formal definitions and remained at the level of philosophical debate. Therefore, this study attempts to formalize agency in a way that explains agent determinism. Our formulation corresponds to one extreme of the spectrum of agency levels and is expected to contribute to clarifying the agency concept.

# Acknowledgement

# Funding

This research was supported by the JSPS KAKENHI (25H00448), Japan. The funding sources had no role in the decision to publish or prepare the manuscript.

# Author Contributions

YO: conceptualization, formalization, draft writing, and revision; AKC: draft writing, and revision; YK: revision.

# Statement and Declarations

The authors declared no potential conflicts of interest with respect to the research, authorship, and/or publication of this article.